
 \input harvmac.tex

\baselineskip=20pt

\Title{Imperial/TP/92-93/29, Princeton PU-PH-93/1392, Swansea SWAT/92-93/5}
{\vbox{\centerline{
Affine Toda Solitons}
\centerline{ and Vertex Operators}}}


\centerline{David I. Olive$^1$, Neil Turok$^2$ and Jonathan W.R. Underwood$^3$}

\centerline{$^1$ Department of Mathematics,
University College of Swansea, Swansea SA2 8PP, Wales, UK.
 }
\centerline{$^2$ Joseph Henry Laboratories,  Princeton University, Princeton,
NJ08544, USA.}

\centerline{$^3$ The Blackett Laboratory, Imperial College, London SW7, UK. }

\centerline{\bf Abstract}

\baselineskip=12pt Affine Toda theories with imaginary couplings
associate with any simple Lie algebra ${\bf g}$ generalisations of
Sine Gordon theory which are likewise integrable and possess soliton
solutions. The solitons are \lq\lq created" by exponentials of
quantities $\hat F^i(z)$ which lie in the untwisted affine Kac-Moody
 algebra ${\bf\hat g}$ and ad-diagonalise the principal Heisenberg
subalgebra. When ${\bf g}$ is simply-laced and highest weight
irreducible representations at level one are considered, $\hat
F^i(z)$ can be expressed as  a vertex operator whose square
vanishes. This nilpotency property is extended to  all highest
weight representations of all affine untwisted Kac-Moody algebras in
the sense that the highest non vanishing power becomes proportional
to the level. As a consequence, the exponential series mentioned
terminates and the soliton solutions have a relatively simple
algebraic expression whose properties can be studied in a general
way. This means that various physical properties of the soliton
solutions can be directly related to  the algebraic structure. For
example, a classical version of Dorey's fusing rule follows from the
 operator product expansion of two $\hat F$'s, at least when ${\bf
g}$ is simply laced. This adds to the list of resemblances of the
solitons with respect to the particles which are the quantum
excitations of the fields.

\Date{April 1993} 


\def\ni{\noindent}

\def\dgh{\Delta({\bf \hat g})}
\def\gh{{\bf \hat g}} \def\g{{\bf g}} \def\dg{\Delta(\g)}
\def\th{\hat\tau}\def\td{\tilde\tau}
\def\em{\hat E_M}\def\en{\hat E_N}\def\ei{\hat E_1}\def\emi{\hat E_{-1}}
\def\fiz{\hat F^i(z)}\def\fio{\hat F^i_0}\def\fjo{\hat F^j_0}
\def\fjz{\hat F^j(\zeta)}
\def\rlj{|\Lambda_j>}\def\llj{<\Lambda_j|}\def\rlo{|\Lambda_0>}
\def\llo{<\Lambda_0|}
\def\gin{\gamma_i\cdot q([N])}\def\gjn{\gamma_j\cdot q([N])}
\def\xij{X_{i,j}(z,\zeta)}
\def\ffiz{\hat{I\!\!F}^i(z)}\def\ffjz{\hat{I\!\!F}^j(\zeta)}
\def\fkz{\hat F^k(\omega^{-q}\zeta)}\def\ffkz{\hat{I\!\!F}^k(\omega^{-q}\zeta)}
\def\gt{{\bf g_{\tau}}}\def\dgt{\Delta(\gt)}
\def\fif{\hat F^{i(1)}(z_1)}\def\fit{\hat F^{i(2)}(z_2)}
\def\ad{\hbox{ad}}\def\epx{e^{-\mu\ei x^+}}\def\emx{e^{-\mu\emi x^-}}
\noindent{\bf 1. Introduction}
\bigskip
This work continues that of our previous paper \ref\OTU{D.I.Olive, N. Turok and
J.W.R.Underwood, Solitons and the Energy-Momentum Tensor for Affine Toda
Theory; Preprint Imperial/TP/91-92/35 and SWAT/3} in which we identified the
manner whereby solutions describing any number of solitons fitted into the more
general class of solutions of the affine Toda theories associated with an
affine
untwisted Kac-Moody algebra $\gh$. Substitution into the energy-momentum tensor
confirmed that the energy was real and positive, being of the form appropriate
to a set of moving, relativistic particles. This was despite the fact that the
solutions  were complex as a consequence of the
coupling constant being imaginary in order to provide topological stability.
Although the energy density was likewise complex, the reality of its integral
was a consequence of the fact that it was  a total derivative which was
asympotically real. This work therefore extended previous results of the last
year
\ref\Holl{T. Hollowood, {\it Nucl. Phys.} {\bf B384} (1992) 523} \ref\LOT{H.C.
Liao, D.I. Olive and N. Turok; {\it Phys. Lett.} {\bf B298} (1993) 95},
\ref\MM{N.J.MacKay and W.A.McGhee; preprint DTP-92-45/RIMS-890}, \ref\CFGZ{C.P.
Constantinidis, L.A. Ferreira, J.F. Gomes and A.H. Zimerman: {\it Phys. Lett.}
{\bf B298} (1993),88}, putting them on a more systematic and general footing
and
unifying them with other developments. Related recent developments are due to
Niedermaier \ref\Nie{M.R. Niedermaier, The Spectrum of the Conserved Charges in
Affine Toda Theories, DESY preprint 92-105} and Aratyn et. al.
\ref\Ar{H.Aratyn,
C.P. Costantinidis, L.A. Ferreira, J.F. Gomes and A.H. Zimerman, Hirota's
Solitons
in the Affine and the Conformal Affine Toda Models, IFT-P.052/92 preprint}.
 \medskip The soliton solutions could be
expressed simply in terms of rational functions of expectation values of the
$r$
\lq\lq fields" $\fiz$, where $r=\hbox{rank }\g$, which diagonalise the
ad-action
of the principal Heisenberg subalgebra of $\gh$. The role of this Heisenberg
subalgebra in other, non-relativistic, soliton theories was first established
by
the Kyoto school \ref\JM{M. Jimbo and T. Miwa, {\it RIMS, Kyoto Univ.} {\bf 19}
(1983) 943}, but the affine Toda theory is of particular interest as it
possesses
relativistic symmetry in space-time with two dimensions and well illustrates
\ref\HM{T. Hollowood and P. Mansfield, {\it Phys. Lett.} {\bf 226B} (1989) 73}
Zamolodchikov's idea \ref\Zam{A.B. Zamolodchikov, {\it Int. J. Mod. Phys.} {\bf
A1} (1989) 4235} that  integrable theories result from  particular breakings of
a
conformally invariant theory in which the conservation laws are relics of the
conformal (or W-) symmetry. Thus it is possible to relate the structure found
in
affine Toda theory to the concepts of particle physics in an intriguing way and
it
is this which motivates our interest in precisely  this theory. For example,
mass
formulae and coupling rules for the solitons emerge bearing remarkable
similarities to those enjoyed by the quantum particle excitations of the
original
fields, as elucidated in recent years \ref\MOP{A.V. Mikhailov, M.A. Olshanetsky
and A.M. Perelomov, {\it Comm. Math. Phys.} {\bf 79} (1981) 473},
\ref\BCDS{H.W.
Braden, E. Corrigan, P.E. Dorey and R. Sasaki, {\it Phys. Lett.} { \bf B227}
(1989) 411; {\it Proc. XVIII Int. Conf. on Differential Geometric methods in
theoretical physics: Physics and geometry, Lake Tahoe,} 1989; {\it Nucl. Phys.}
{\bf B338} (1990) p465.}, \ref\CM{P. Christe and G. Mussardo, {\it Nucl. Phys.}
{\bf B330} (1990) 465; {\it Int. J. Mod Phys.} {\bf A5} (1990) 4581},
 \ref\Dor{P.E. Dorey,
{\it Nucl. Phys.} {\bf B358} (1991) 654},
\ref\Fre{M.D. Freeman, {Phys. Lett.} {\bf B261} (1991) 57 }, \ref\FLO{A. Fring,
H.C. Liao and D.I. Olive, {\it Phys. Lett.} {\bf B 266} (1991) 82. }. This
points
to a duality symmetry in which the inversion of the coupling constant is
accompanied by a replacement of the roots of $\g$ by the corresponding coroots.
All this structure is very reminiscent of ideas entertained fifteen years ago
in
connection with magnetic monopole solitons arising in spontaneously broken
gauge
theories in four dimensional space time \ref\GNO{P. Goddard, J. Nuyts and
D.I.Olive, {\it Nucl. Phys.} {\bf B125} (1977) 1. }, \ref\O{D.I. Olive, \lq\lq
Magnetic Monopoles and Electromagnetic Duality Conjectures" in \lq\lq Monopoles
in
Quantum field Theory", edited by N.S. Craigie, P. Goddard and W. Nahm (World
Scientific 1982) p157-191}, \ref\KW{H. Kausch and G.M.T. Watts,  {\it Nucl.
Phys.} {\bf B386} (1992) 166. }, \ref\Holb{T. Hollowood, {\it Int. Journ. Mod.
Phys.} {\bf A8} (1993)
947-981. }. The difference here is that the
construction of the dual quantum field theory describing the solitons is now a
more
realistic possibility.

\medskip
Our construction of the soliton solutions has the intuitively
attractive feature that the soliton of species \lq\lq$i$" is \lq\lq
created" by the \lq\lq Kac-Moody" group element given by the exponential
$$exp\,
Q\fiz,\eqno(1.1)$$ where $ln\,|Q|$ and $ln\,|z|$ relate to the coordinate and
rapidity of the soliton respectively while the superscript \lq\lq $i$",
denoting
the soliton species, labels one of the $r$ orbits of the $hr$ roots of $\g$
under the action of the Coxeter element of its Weyl group. It will be
convenient to take as the representative element of each such orbit the unique
one
of the form $$\gamma_{(i)}=c(i)\alpha_{(i)}\eqno(1.2)$$ where
$\alpha_{(i)}$ is simple and $c(i)=\pm1$ denotes its \lq\lq colour", black or
white, as explained in \Dor and \FLO. That the exponentials (1.1) are
meaningful as
Kac-Moody group elements without any normal ordering is a consequence of the
fact
that $\fiz$ is nilpotent: in an irreducible representation of $\gh$  at level
$x$,
powers of $\fiz$ higher than $2x/\gamma_{(i)}^2$ vanish while lower powers
usually
have finite matrix elements. (In saying this we have chosen to normalise the
highest
root of $\g$ to have length $\sqrt2$). (The exceptions to this finiteness will
have a physical interpretation to be discussed beow). \medskip
The present paper supplies the proof of these properties. At
the same time techniques are found for calculating the expectation
values of the $F$'s needed to evaluate the soliton solutions
explicitly. The techniques used have an intrinsic interest; the
principal vertex operator construction is used together with an exploitation of
the outer automorphisms
of $\gh$ corresponding to symmetries of the extended Dynkin diagram of
$\g$. By-products of interest to the physical interpretation of the
soliton solutions concern (a) a classical analogue of Dorey's fusing
rule \Dor applicable to soliton solutions rather than the quantum
excitation particles, (b) insight into generalised \lq\lq breather"
solutions, and (c) expressions for the weight lattice coset of the
topological charge when $g$ is simply laced.
\medskip
In section 2 we present our notations and recapitulate the equations
considered together with our solutions. In section 3 we establish properties of
the outer automorphisms of $\gh$ to be used while in section 4 we present the
properties
of the expectation values of the $F$'s with respect to the highest
weight states of the fundamental representations of $\gh$. These form a
pseudo-unitary matrix which plays an important role in subsequent
work.
\medskip
Since the results sought are representation dependent we have to
proceed by considering successively more elaborate irreducible
representations of $\gh$. The simplest possibility occurs when $\g$ is
simply laced and at level one, being known as the \lq\lq basic"
representation. As the $\g$ weight of its highest weight state
vanishes, that state can be thought of as the vacuum. In this
representation, the $F$'s can be constructed explicitly in terms of
the principal Heisenberg subalgebra via the principal vertex operator
construction \ref\KKLW{V.G. Kac, D.A. Kazhdan, J. Lepowsky,  and R.L. Wilson,
{\it Advances
in Math.}  {\bf 42} (1981) 83-102. }, \ref\Kac{V.G.  Kac, \lq\lq
Infinite-dimensional Lie
algebras" , Third Edition, {\it Cambridge University Press}, 1990} as explained
in section 5. It
is therefore possible to calculate the operator product expansion of two $F$'s.
The result
possesses double poles with c-number residues and simple poles whose residues
are proportional to
elements of the principal Heisenberg subalgebra or a third $F$. The
$F$'s so occurring satisfy Dorey's fusing rule and this is the
algebraic origin of this rule as reported previously \OTU . It is easy
to check that the square of each $F$ vanishes and to \lq\lq normal order"
products of $F$'s in order to calculate matrix elements, rather as is done in
evaluating dual string scattering amplitudes. Section 6 extends
these results to the other irreducible representations of $\gh$ at levels
higher than one (when $g$ is simply laced) in which the vertex operator
construction no longer
applies. \medskip
Section 7 treats non-simply laced Lie algebras, denoted $g_{\tau}$,
as they are obtained in a unique way by the familiar folding procedure from a
simply laced $\g$ and an outer automorphism $\th$ \ref\OTa{D.I Olive and N.
Turok,
{\it Nucl. Phys.}  {\bf B220} [FS8] (1983) 491-507.
 }, \ref\GNOS{ P. Goddard, W. Nahm, D.I. Olive and A. Schwimmer, {\it Comm.
Math. Phys.}
 {\bf 107} (1986) 179-212.}. The key
result is that $\th$ preserves the principal $su(2)$ subalgebra of
$\g$. As a consequence, the principal Heisenberg subalgebra of
$\g_{\tau}$ is a subalgebra of that of $\g$. This makes it   easy to relate
the $F$'s for $\gh$  and $\gh_{\tau}$ and hence establish the final versions of
the quoted nilpotency and operator product properties. \medskip
The physical interpretation of these results, (a), (b) and (c), cited
four paragraphs above, are
discussed in section 8. Sample calculations of soliton solutions
are also given for $su(N)$, $so(8)$ and $G_2$.
\bigskip
\noindent{\bf 2. The Affine Toda Equations and their Soliton Solutions}
\bigskip

even though it is not difficult to generalise virtually all of our arguments to
the latter. Before
presenting the equations and the solutions of interest, we shall summarise our
conventions for such algebras and their representations.
\medskip
\ni{\it 2.1 The Chevalley Basis and the Principal Gradation}
\medskip

The Chevalley basis of the affine untwisted Kac-Moody algebra $\gh$
is generated by the elements $$\{e_i,f_i,h_i\}\qquad i=0,1\dots r.$$
It has been proven \Kac that $$\eqalign{[h_i,h_j]=&0\cr
[h_i,e_j]=&K_{ji}e_j\cr [h_i,f_j]=&-K_{ji}f_j\cr
[e_i,f_j]=&\delta_{ij}h_i\cr}\eqno(2.1)$$ together with the Serre
relations $$\eqalign{(\ad e_i)^{1-K_{ji}}e_j=&0\cr (\ad
f_i)^{1-K_{ji}}f_j=&0\cr}\eqno(2.2)$$ are sufficient to completely
specify the algebra, given  $K_{ji}$, the Cartan matrix of $\gh$,
satisfying the usual properties. This is  the analogue of Serre's
theorem for a finite-dimensional Lie algebra. It is then convenient
to add the element $d'$ with the defining properties
$$[d',e_i]=e_i,\qquad[d',h_i]=0\qquad\hbox{ and }\qquad
[d',f_i]=-f_i, \eqno(2.3)$$ and to regard it, together with the
$h_i$, as spanning the Cartan subalgebra which therefore has
dimension $r+2$. Notice that the adjoint action of $d'$ grades
$\gh$; this is generally known as the principal gradation in view of
its connection with the principal $so(3)$ subalgebra of $\g$ as seen
below. Since the $e_i$ can be regarded as the step operators for the
simple roots of $\gh$, the grade of a step operator  counted by $d'$
is simply the height of the associated root. \medskip We define the
positive integers $m_i$ and $n_i$ to be the lowest for which
$$\sum_iK_{ji}m_i=0\qquad\hbox{ and
}\qquad\sum_jn_jK_{ji}=0.\eqno(2.4a)$$ It follows
that$$n_i=2m_i/a_i^2\eqno(2.4b)$$where $a_i$ denotes the $i$'th
simple root of $\gh$ and the long roots are chosen to have length
$\sqrt2$. The Coxeter and dual Coxeter numbers of $\g$ are defined
respectively as $$h=\sum_{i=0}^rn_i\qquad\hbox{ and }\qquad\tilde
h=\sum_{i=0}^rm_i.\eqno(2.5)$$ The following element of the Cartan
subalgebra is central in that it commutes with all of $\gh$:-
$$x=\sum_{i=0}^rm_ih_i.\eqno(2.6)$$  We shall want to compare this
presentation of $\gh$, which is particularly relevant to the affine
Toda theory, to the more familiar one as a central extension of the
loop algebra of $\g$: $$[\lambda^m\otimes\gamma_1,
\lambda^n\otimes\gamma_2]=\lambda^{m+n}
\otimes[\gamma_1,\gamma_2]+\delta_{m+n,0}(\gamma_1,\gamma_2)mx,\eqno(2.7)$$
where $\gamma_1,\gamma_2\in\g$ and ( , ) is the Killing form on $\g$.
When $\gh$ is viewed this way it is more natural to append an element
$d$ with the grading property
$$[d,\lambda^m\otimes\gamma]=m\lambda^m\otimes\gamma.$$
This is known as the homogeneous grading and $d$ can be thought of as
the Virasoro generator $-L_0$.
\medskip
Relating the two presentations of $\gh$ is straightforward. For $i\neq0$
$$e_i\Leftrightarrow\lambda^0\otimes E^{\alpha_i},\qquad
f_i\Leftrightarrow\lambda^0\otimes E^{-\alpha_i},\qquad
h_i\Leftrightarrow\lambda^0\otimes H^{\alpha_i},\eqno(2.8a)$$
in terms of the Chevalley generators of $\g$ while, for $i=0$,
$$e_0\Leftrightarrow \lambda^1\otimes E^{-\psi},\qquad
f_0\Leftrightarrow\lambda^{-1}\otimes E^{\psi}, \qquad
h_0\Leftrightarrow\lambda^0\otimes (H^{-\psi}+x),\eqno(2.8b)$$
where $\psi\equiv-\alpha_0$ denotes the highest root of $\g$. Finally
we can now state the relation between $d$ and $d'$ which count the
homogeneous and principal grades respectively
$$d'=hd +T_0^3.\eqno(2.9)$$
Here $T_0^3$ is defined to be $\lambda^0\otimes T^3$ where $T^3$ is the
generator of the principal or maximal $so(3)$ subalgebra of $\g$ lying
in the Cartan subalgebra of the latter. Its adjoint action on the step
operators of $\g$
counts the height of the corresponding roots. The analogue of $d'$ for
the finite dimensional Lie algebra  $\g$ must be obtained by somehow
modding out $d$. By (2.9), this is achieved by defining
 the element of the corresponding Lie group $G$, $S=e^{2\pi iT^3/h}$.
 Conjugation with respect to
$S$ then
 grades $\g$ according to powers of $\omega=exp(2\pi i/h)$. This
structure played a crucial role in understanding the properties of the
affine Toda particles arising as field quanta \Fre, \FLO, and is explained
further in section 2.4.
    \bigskip
\ni{\it 2.2 Highest Weight Representations}
\medskip
The construction of solutions to the affine Toda equations will
utilise irreducible highest weight representations of $\gh$, particularly the
fundamental ones. The corresponding representation spaces are
generated by the action of arbitrary products of the $f_i$ on the
\lq\lq highest weight" state $|\Lambda>$, say, which is annihilated by the
$e_i$.
The representation is precisely characterised by the action of $h_0,
h_1\dots h_r$ on this state, which must be an eigenvector under this
action. We write
$$h_i|\Lambda>=\Lambda(h_i)|\Lambda>={2\Lambda\cdot a_i\over
a_i^2}|\Lambda>.\eqno(2.10)$$
The scalar product here is deduced from the invariant scalar product on
$\gh$. Unitarity requires the $r+1$
quantities
 $\Lambda(h_i)$ to be
non-negative integers. It follows that the eigenvalue of $x$ take the
same positive integral value on all states of the irreducible
representation, which we denote $V(\Lambda)$. This is known as the
level,
 and is again denoted $x$,
by abuse of notation. The number of irreducible
representations at each level  is finite and at least one.
\medskip
It is convenient to define the \lq\lq fundamental" weights
$\Lambda_0,\Lambda_1,\dots \Lambda_r$ of $\gh$ by
$\Lambda_i(h_j)=\delta_{ij}$. Since the $r+1$ simple roots $a_i$ of $\gh$
relate to the $r$ simple roots $\alpha_i$ of $\g$ by
$$a_0=(-\psi,1,0),\qquad a_i=(\alpha_i,0,0)\quad i=1,2,\dots r,\eqno(2.11)$$
given the Cartan Weyl basis of the Cartan subalgebra of $\gh$,
$(H_0^i,x,d)$, we find
$$\Lambda_0=(0,1,?)\qquad \Lambda_i=(\lambda_i,m_i,?)\quad i=1,2,\dots
r.\eqno(2.12)$$
The entries denoted by the question mark are undetermined since the
roots span a subspace of codimension $1$ but are conventionally taken
to vanish. The levels $m_i$ in (2.12) are the integers defined in
(2.4). For more details, see \ref\GO{P. Goddard, and D.I. Olive, {\it Int. J.
Mod. Phys }{\bf A1} (1986) 303. } where the same notation is used.
\medskip
Notice that the state $\otimes_j\rlj^{p_j}$ of the representation
$\otimes_jV(\Lambda_j)^{p_j}$, where the $p_j$ are positive integers,
is automatically annihilated by the $f_i$ and so is a highest weight
state. Thus $$\otimes_j\rlj^{p_j}=|\sum_jp_j\Lambda_j>,\eqno(2.13)$$ where the
right hand side denotes the highest weight state  of
$V(\sum_jp_j\Lambda_j>$. Thus, in principle, all irreducible
representations of $\gh$ can be found by the decomposition of products
of its fundamental representations. In fact it is possible to improve on
this by considering products of only those fundamental representations
with level one, ie $m_i=1$. These include the vacuum representation, that
with highest weight $\Lambda_0$. If $\g$ is simply laced this condition means
that $\lambda$ either vanishes or is minimal \GO,\ref\Humphreys{J.E. Humphreys,
Introduction to Lie Algebras and Representation Theory, (Springer, Heidelberg,
1972)}
 \bigskip \ni{\it 2.3 Alternative basis for $\gh$}
\medskip
The subspace of $\gh$ with principal, $d'$ grade unity contains an
element of special significance
$$\hat E_1=\sum_{i=0}^r\sqrt{m_i}\,\,e_i.\eqno(2.14)$$
As we see in section 3 it is invariant under all the diagram
automorphisms of $\gh$. It was found some time ago \ref\LS{A.N. Leznov,  and
M.V.
Saveliev, {\it Lett. in Math. Phys.} {\bf 3} (1979) 485.   } that the
integrability of the affine Toda equations owed itself to the
existence of field dependent zero curvature potentials \ref\Bog{O.I.
Bogoyavlensky, {\it Commun. Math. Phys.} {\bf 51} (1976), 201; A.N. Leznov and
M.V. Saveliev, {\it Lett. Math. Phys.} {\bf 3} (1979) 489; M.A. Olshanetsky and
A.M. Perelomov, {\it Invent. Math.} {\bf 54} (1979) 261}. These can be lifted
from
the loop algebra of $\g$ to $\gh$ at the cost of introducing an extra,
innocuous,
field \ref\OTc{D.I.Olive and N. Turok, {\it Nucl.Phys.} {\bf 265B [FS15]}
(1986)
469.}. The constant, vacuum solutions to the affine Toda equations give rise to
potentials depending linearly on $\hat E_1$, (2.14), and its conjugate,
 $\hat E_{-1}=\hat E_1^{\dagger}$. We
see from (2.1) and (2.7) that
$$[\hat E_1,\hat E_{-1}]=x.$$
The affine Toda Hamiltonian is one member of an infinite hierarchy of
integrable systems and consideration of this makes it natural to seek
the subalgebra $C(\hat E_1)$ that commutes with $\hat E_1$, modulo the
central term $x$. It is not difficult to show that can be expressed as
a direct sum of graded subspaces
$$C(\hat E_1)=\oplus\sum_{M\in Z\!\!\!Z}C_M(\hat E_1) \quad\hbox{
where }[d',C(\hat E_1)]=MC_M(\hat E_1).$$
The integers $M$ for which dim $C_M(\hat E_1)$ is non-zero are called
exponents of $\gh$ of multiplicity dim $C_M(\hat E_1)$. Tables of
these exponents can be found in \Kac. These exponents possess the
symmetry $M\leftrightarrow-M$ and possess period $h$, the Coxeter
number of $\g$. Modulo $h$ the exponents of $\gh$ equal the exponents
of $\g$.  Since the only occasion in which the
multiplicity exceed one (and equals two) is when $\g=D_{2n}$ and the
exponent equals $n-1 $ mod $2n-2$, we shall permit ourselves the
temporary simplifying assumption that $C_M(\hat E_1)$ is spanned by
$\em$ when $M$ is an exponent.
Using the invariant bilinear form ( , ) on $\gh$ it follows that \OTU
$$[\em,\en]={Mx\over h}(\em,\en)\delta_{M+N,0},$$
So the $\em$ generate what is known as the principal Heisenberg subalgebra of
$\gh$. We suppose $\em^{\dagger}=\hat E_{-M}$ and choose to normalise
consistently with (2.14) so that
$$[\em,\en]=Mx\delta_{M+N,0}.\eqno(2.15)$$
\medskip
It is a remarkable fact that, unless $M=0$, the dimension of the
principally $M$-graded subspace $\gh_M$ of $\gh$ equals $r+$ dim
$C_M(\hat E_1)$. The remaining $r$ dimensions of the $M$ graded
subspace are spanned by vectors $\hat F_M^1,\hat F_M^2,\dots \hat F_M^r$
obeying the following commutators with the Heisenberg subalgebra
\OTU
$$[\em,\hat F_N^i]=\gamma_i\cdot q([M])\hat F_{M+N}^i.\eqno(2.16)$$ It is
understood that $[M]$ denotes $M$ modulo $h$ and that $q([M])$ and $\gamma_i$
lie
in the root space of $\g$. These quantities arise naturally in the construction
of
the Cartan subalgebra of $\g$ in apposition, a structure which is essentially
the
analogue for $\g$ of the construction above for $\gh$ as we
briefly explain in the next subsection. (More details can be found in Chapter
14
of \Kac
and in  appendix A of \OTU.)
\medskip
If we define the generating function for the $\hat F_N^i$ in terms of
a formal complex parameter $z$  as
$$\fiz=\sum_{N=-\infty}^{\infty}\, z^{-n}\hat F_N^i,\eqno(2.17)$$
we find from (2.12) that
$$[\em,\fiz]=\gamma_i\cdot q([M])z^M\fiz.\eqno(2.18)$$
Thus the complete Heisenberg subalgebra of $\gh$ has been
ad-diagonalised by the $\fiz$, which, as we mentioned in the
introduction, create the solitons of affine Toda theory when
exponentiated, (1.1).
\bigskip
\ni{\it 2.4 A construction of the Alternative Basis for $\gh$}
\medskip
The construction outlined above has  a well established analogue for
the finite-dimensional Lie algebra $\g$. Indeed this played a
fundamental role in the construction of the local conservation laws of
affine Toda theory and in understanding the mass and coupling
properties of the particles which are the quantum excitations of the
$r$ fields  \Dor, \Fre, \FLO. The mathematics of the construction is due to
Kostant \ref\Kos{B. Kostant,{\it Amer. J. Math.} {\bf 81} (1959) p973}. The
analogue of (2.14) in $\g$ is
$$E^1=\sum_{i=1}^r\sqrt{m_i}\,E^{\alpha_i}+E^{-\psi}.\eqno(2.19)$$ Now it
commutes
with its conjugate and, in fact, as $E^1$ is in \lq\lq general position", the
subalgebra of $\g$ commuting with it possesses $r$ dimensions and so is
abelian.
It can therefore be considered as an alternative Cartan subalgebra ${\bf h'}$,
say, said to be \lq\lq in apposition". Conjugation by $S=e^{2\pi iT^3/h}$
furnishes a linear map $\sigma$ on ${\bf h'}$. $\sigma$ can be shown to be an
element of the Weyl group, of order $h$, called the Coxeter element. This
possesses $r$ eigenvalues of the form $e^{2\pi i\nu/h}$, where $1\leq\nu\leq
h-1$
is one of the exponents of $\g$. Accordingly, if $h_1,h_2,\dots h_r$ denotes an
orthonormal basis of ${\bf h'}$, we can express the $S$-graded elements of
${\bf
h'}$ as $$SE^{\nu}S^{-1}=e^{{2\pi i\nu\over h}}S, \qquad E^{\nu}=q(\nu)\cdot
h\qquad\hbox{where}\qquad\sigma\left(q(\nu)\right)=e^{{2\pi i\nu\over
h}}q(\nu),\eqno(2.20)$$
so that the $q(\nu)$ are the eigenvectors of $\sigma$. Because $\sigma$
is unitary and real we can orthonormalise in the following sense
$$q(\nu)\cdot q(\nu')^*=\delta_{\nu,\nu'}, \qquad
q(h-\nu)=q(\nu)^*.\eqno(2.21)$$
This is explained in more detail in \FLO. There it is also explained
how it is possible to define the step operators $F^{\alpha}$ for the roots with
respect to ${\bf h'}$ and that $\sigma$ splits them into precisely $r$
orbits each containing $h$ roots. From the definition (2.20),
we have the commutator
$$[E^{\nu},F^{\alpha}]=\alpha\cdot q(\nu)F^{\alpha}.\eqno(2.22)$$
The modulus of the structure constant here is proportional to the mass of the
affine Toda particles. The fact that this is also proportional to the right
Perron Frobenius eigenvector $y_j(1)$ of the Cartan matrix of $\g$
  follows from the important identity
$$\gamma_j\cdot q(\nu)=2i\sin\,{\pi\nu\over h}y_j(\nu)e^{-i\delta_{jB}\pi
i/\nu}.\eqno(2.23) $$
where $\delta_{jB}=1$ if the vertex $j$ is \lq\lq black", and zero
otherwise. $\gamma_j$ is defined in (1.2), being the standard representative of
the $i$-th orbit.
 \medskip
It is possible to affinise $\g$ to form $\gh$ following the
construction (2.7) but using the basis just discussed. The resulting
quantities $E_m^{\mu}$ and $F_m^{\alpha}$ are graded with respect to
$d$, the homogeneous grade, rather than $d'$, the principal grade.
Regarding $z$ as a formal parameter, the following formula succinctly
converts between the grades
$$\sum_{M\in Z\!\!\!Z}z^{-M}\em=\sum_{m\in Z\!\!\!Z}z^{-mh}z^{-T_0^3}
\left(\sum_{\nu}E_n^{\nu}\right)z^{T_0^3}.\eqno(2.24)$$
This is equivalent to the result derived in the appendix of \OTc. (See also
\ref\KP{V.G. Kac and D.
Peterson, \lq\lq 112 Constructions of the Basic Representation of the Loop
Group of $E_8$", Symposium
on Anomalies, Geometry and Topology, edited by W.A. Bardeen and A.R. White,
World Scientific,
Singapore, 1985.} )

 A similar expression applies to the $\fiz$
$$\fiz=\sum_Mz^{-M}\hat
F_M^i=\sum_{m\in Z\!\!\!Z}z^{-mh}z^{-T_0^3}F_m^{\gamma_i}z^{T_0^3}
-x(T^3,F^{\gamma_i})/h.\eqno(2.25)$$
The necessity for the last, central, term was explained in \OTU.

\bigskip

\ni{\it 2.5 The Affine Toda Equations and their Soliton Solutions}
\medskip

The affine Toda field theory equations associated with $\gh$ comprise
a set of $r$ coupled, differential equations satisfied by $r$ scalar
fields $\phi$, taken to lie in the root space of $\g$, taking the
relativistically invariant form in two space-time dimensions
$$\partial^2\phi+{4\mu^2\over\beta}\left(\sum_{i=
1}^rm_i{\alpha_i\over\alpha_i^2}e^{\beta\alpha_i\cdot\phi}
-{\psi\over\psi^2}e^{-\beta\psi\cdot\phi}\right)=0.\eqno(2.26)$$
$\mu$ denotes  a real inverse length scale and $\beta$ an imaginary
coupling constant. The coefficients are so arranged that $\phi=0$ is a
constant solution. The most general solution was found in a formal
sense in \OTU but we shall be interested in what we claimed to be the
solitonic specialisation of this solution. These specialisations are
given by
$$e^{-\beta\lambda_j\cdot\phi}={\llj e^{-\mu\hat
E_1x^+}g(0)e^{-\mu\hat
E_{-1}x^-}\rlj\over<\Lambda_0|e^{-\mu\hat
E_1x^+}g(o)e^{-\mu\hat
E_{-1}x^-}|\Lambda_0>^{m_j}},\eqno(2.27)$$
for $j=1,2,\dots r$. $g(0)$ was a Kac-Moody group element arising as
an integration constant. The choice which described $n$ solitons was a
product of $n$ factors of the form (1.1), one for each soliton, with
the moduli of the variables $z$ suitably ordered. The \lq\lq time
development" operators could be eliminated, using the identity valid for
$j=0,1,2,\dots r$,
$$\llj e^{-\mu\hat E_1x^+}g(0)e^{-\mu\hat E_{-1}x^-}\rlj=
e^{\mu^2m_jx^+x^-}\llj g(t)\rlj,\eqno(2.28)$$
 where $g(t)$ is obtained by multiplying each occurrence of $\fiz$ by
the factor
 $$W(i,z,x^{\mu})=exp\left(-\mu\{\gamma_i\cdot q(1)zx^+-\gamma_i\cdot
q(1)^*z^{-1}x^-\}\right),\eqno(2.29a)$$
$$=exp\left(\pm\mu|\gamma_i\cdot
q(1)|\{e^{\eta}x^+-e^{-\eta}x^-\}\right),\eqno(2.29b)$$  choosing
 $$z=\mp e^{\eta}|\gamma_i\cdot q(1)|/\gamma_i\cdot
q(1)\eqno(2.30)$$
 for
the reasons explained in \OTU. There $\eta$ was shown to be the
rapidity of the soliton created by this $\fiz$. The result (2.29)
 is a straightforward
application of the commutation relations (2.15) and (2.18).
\medskip
The justification of the specialisation  was that it was possible to evaluate
the energy and
momentum of these solutions, finding the  expressions appropriate to
$n$ moving solitons, given the choice of the variables $z$ explained
in \OTU, despite the complexity implied by the imaginary coupling constant
$\beta$. The remainder of this paper will be devoted to the study of these
formulae. It was crucial in the arguments of \OTU that products of factors were
bona-fide Kac-Moody group elements. This will be justified by developing the
properties of the $\fiz$, some of which are Lie algebraic and some of which,
such
as the nilpotency, are representation dependent.
\bigskip

\ni{\bf 3. Automorphisms of $\gh$ and symmetries of its Dynkin diagram}
\medskip
Let $\dgh$ denote the Dynkin diagram of the affine untwisted Kac-Moody algebra
$\gh$. It is well known that this is the same as the
 extended Dynkin diagram of the finite-dimensional Lie algebra $\g$. $\dgh$
tends to be more symmetrical than $\dg$, the Dynkin diagram of $\g$, and this
will be important in
what follows as it will mean that $\gh$ has more outer automorphisms than $\g$
has.
There will be several ways in which  these can be exploited to simplify
calculations
of matrix elements of the $F$'s and hence the soliton solutions themselves. As
the discussion of this section is rather technical, the reader may wish, at a
first
reading, to pass immediately to section 4.
\medskip The symmetries of $\dgh$ form a finite
group denoted $Aut\dgh$ whose elements can be denoted by permutations of the
vertices of $\dgh$ preserving its structure: $$\tau:i\rightarrow\tau(i)\qquad
\tau\in Aut\dgh.\eqno(3.1)$$ We immediately have a representation of $aut\dgh$
of
dimension $r+1$: $$D_{ij}(\tau)=\delta_{i\tau(j)},\eqno(3.2)$$ where $D$ is a
real, orthogonal matrix. Its character is simply given by $$\chi(\tau)\equiv
Tr(D(\tau))=\hbox{ No. of points of $\dgh$ fixed by } \tau.\eqno(3.3)$$
Given the character table of $Aut\dgh$, (3.3) can be used to deduce the
decomposition of $D$ into irreducible representations. Because of its defining
property, $D(\tau)$ must commute with the Cartan matrix of $\gh$, denoted $K$:
$$[K,D(\tau)]=0.\eqno(3.4a)$$
It also follows from the definitions (2.4) that
$$m_{\tau(i)}=m_i\qquad n_{\tau(i)}=n_i\eqno(3.4b)$$
There is a standard way of lifting any non-trivial $\tau\in Aut\dgh$ to an
automorphism of $\gh$ which is  outer. The key point is that, if $X_i$ denotes
$e_i$, $f_i$ or $h_i$, the map $\th$ defined by
$$X_i\rightarrow X_{\tau(i)}=X_jD_{ji}(\tau)\equiv \th(X_i)\eqno(3.5)$$
respects the defining relations (2.1) and (2.2) of $\gh$ by virtue of (3.4) and
hence can
be extended to an automorphism of $\gh$, by the reconstruction theorem. Notice
that, by (2.6), (3.4) and (3.5), the level, $x$, is invariant
$$\th(x)=x.\eqno(3.6)$$

 \medskip
$\th$ can readily extended to the enveloping algebra by imposing
$\th(ab)=\th(a)\th(b)$ and similarly to states of representations of $\gh$. If
$|\Lambda>$ denotes a highest weight state, so that it is annihilated by all
$r+1$ $e_i$, then, by (3.5), so is $\tau(|\Lambda>)$.  Furthermore, if it is
one
of the $r+1$ fundamental highest weight states so that
$$h_i|\Lambda_j>=\delta_{ij}|\Lambda_j>,$$
we find, by (3.5) that
$$\th|\Lambda_i>\sim |\Lambda_{\tau(i)}>.\eqno(3.7)$$
As long as $\tau$ is non-trivial there exists an $i$ for which
$\tau(i)$ is distinct from it. Thus as $|\Lambda_i>$ and $|\Lambda_{\tau(i)}>$
are highest weight states of inequivalent irreducible representations
of $\gh$, we conclude that $\th$ is an outer automorphism of $\gh$.
In fact it is known that the group of outer automorphisms of $\gh$,
understood as the quotient of the group of all automorphisms by the
invariant subgroup of inner automorphisms is a finite group isomorphic
to $Aut\dgh$.
\medskip Because of the way the soliton solutions are expressed in
terms of the principal Heisenberg subalgebra $\{\em\}$ and the $F$'s
which ad-diagonalise it, we need to determine the action of $\th$ on
these quantities. The first comment is that, by (3.5), $\th$ leaves
invariant the
subspaces of $\gh$ with principal grade $1$, $-1$ and $0$
respectively. As these subspaces generate all of $\gh$ it follows
that $\th$ always respects the principal grade, which is counted by
the adjoint action of $d'$. Furthermore, by its explicit form, (2.21),  $\hat
E_1$ is invariant:
$$\th(\hat E_1)=\hat E_1 \qquad\hbox{ for all }\tau\,\in\, Aut\dgh.\eqno(3.8)$$
Since the $\em$ are characterised by their principal grade and their
commutation relation with $\hat E_1$, we conclude that
$$\th(\em)=\eta(\tau,M)\em, \qquad M=\hbox{exponent of }\gh,\eqno(3.9)$$where
$\eta(\tau,M)$ denotes a two dimensional representation of $Aut\dgh$ when
$\g=D_{2k}$ and $M=2k-1$ mod $4k-2$ and a one dimensional representation
otherwise (that is, a phase). We shall need explicit formulae for $\eta$ and
shall
calculate it in all cases of interest. Most often it simply equals unity, as in
(3.8),
but there are crucial cases when it does not.
\medskip
To proceed further, we need to know more about the structure of $Aut\dgh$ and
this can be deduced by considering its action on $\g$ rather than $\gh$.
That it has such an action follows from the comment that $\dgh$ is also the
extended Dynkin diagram of $\g$ defined by the $r$ simple roots augmented by
the negative of the highest root. Therefore
$$Aut\dgh\subset Aut\Phi(\g),$$
where $\Phi(\g)$ denotes the root system of $\g$. The structure of
$Aut\Phi(\g)$ is well understood \Humphreys. If $W(\g)$ denotes the Weyl group
of $\g$, it constitutes an invariant subgroup of $Aut\Phi(\g)$ such that
$$Aut\Phi(\g)/W(\g)\cong Aut\dg,$$
where $Aut\dg$ itself can be thought of as the subgroup of $Aut\dgh$ respecting
the vertex labelled $0$. Let us define
$$W_0(\g)=Aut\dgh\cap W(\g),\eqno(3.10)$$
and analyse its  structure. We see immediately that $W_0$ is likewise an
invariant subgroup of $Aut\dgh$ and that,
$$Aut\dgh/W_0(\g)\cong Aut\dg.\eqno(3.11)$$
We can now deduce that any $\tau\,\in\, Aut\dgh$ has a unique decomposition
$$\tau=\rho\tau_0,\quad \rho\,\in\, Aut\dg,\quad \tau_0\in
W_0.\eqno(3.12)$$
 By (3.11) there exists $\tau_0\,\in\, W_0$ with
$\tau_0(0)=\tau(0)$. Hence $\tau\tau_0^{-1}=\rho\,\in\, Aut\dg$. Furthermore,
$\tau_0$ is unique, for, if not, and $\tau_0(0)=\tau_0'(0)$, we would have
$\tau_0'\tau_0^{-1}\,\in\, W_0\cap Aut\dg$. As this intersection contains only
the unit element the result (3.12) follows.
\medskip We conclude that the elements of
$W_0$ are labelled by the \lq\lq tip points" of $\dgh$, namely all those
vertices symmetrically related to the vertex labelled $0$. The fundamental
$\g$-weights associated to the tip points of $\dgh$ other than $0$ itself are
all minimal \Humphreys, and hence (when $0$ is included) are in one-to-one
correspondence with the cosets of the weight lattice of $\g$ with respect to
its
root lattice, and hence with the centre, $Z(\g)$, of the simply connected Lie
group whose Lie algebra is $\g$. This suggests that $W_0(\g)$ and $Z(\g)$ are
related. In fact it has been checked that they are isomorphic \OTa. We can make
the isomorphism more explicit once we lift $\tau$ in $Aut\dgh$ to a
automorphism
$\td$ of the finite dimensional Lie algebra  $\g$.
\medskip
The automorphism (3.5) of $\gh$ worked at any level so, in particular, we may
choose level zero in which case $\gh$ is realised by the loop algebra
$$e_i=E^{\alpha_i}\quad i=1,\dots r\qquad e_0=\lambda E^{\alpha_0},$$
where we have written the step operators for the extended set of roots of $\g$.
With similar expressions for the $f_i$, we see that (3.5) now furnishes an
automorphism of $\g$ rather than $\gh$. We denote this $\td$ in order to
distinguish it from $\th$. The difference is that now $\tau$ in $W_0$ furnishes
an inner automorphism of $\g$. We may as well put $\lambda$ equal to unity. We
see that  $$\eta (\tau,N)=\eta(\tau,\nu),\eqno(3.13)$$
where $N=nh(g)+\nu$, and $\nu$ is an exponent of $\g$ and so taking values
between $1$ and $h-1$. Furthermore, by the construction of $E_{\nu}$ in the
appendix of \ref\OTb{D.I. Olive and N.Turok, {\it Nucl. Phys.} {\bf 257B
[FS14] }(1985) 277-301. }  ,
$$\eta(\tau,\nu)=1\qquad \tau\in W_0.\eqno(3.14)$$
Furthermore $\eta$ only depends on $\tau$ through the coset (3.11). We can now
apply $\td$ to $S=exp(2\pi iT^3/h)$, where $T^3=\rho^v\cdot
H=\sum_{\alpha>0}2\alpha\cdot H/\alpha^2$ is the intersection of the principal
$so(3)$ with its Cartan subalgebra. Taking scalar products with the simple
roots of $\g$ in turn enables us to deduce
$$\tau(\rho^v)-\rho^v=-{2h\lambda_{\tau(0)}\over\alpha_{\tau(0)^2}},$$
from which we find
 $$\td(S)=S \,exp(-4\pi i\lambda_{\tau(0)}\cdot
H/\alpha_{\tau(0)}^2).\eqno(3.15)$$ As the second factor lies in $Z(\g)$ this
establishes the isomorphism $$W_0(\g)\cong Z(\g).\eqno(3.16)$$ This establishes
the fact that $W_0$ is abelian, and, in fact, it is maximally so in $\dgh$ as
any element therein, commuting with all elements of $W_0$, must lie in $W_0$.
We
shall see how (3.16) affords important information concerning the asymptotic
behaviour of soliton solutions in section (8.1). \medskip Applying (3.9) and
(3.14)
to the statement that $\fiz$ ad-diagonalises the $\em$, we deduce that
$$\th(\fiz)=\epsilon(\tau,i)\fiz,\qquad \tau\in W_0(\g),\eqno(3.17)$$
where $\epsilon(\tau,i)$ is one of the irreducible (and hence one-dimensionsal)
representations of $W_0\cong Z$. In sections 4 and 8 we shall present arguments
to the effect that
$$\epsilon(\tau,i)=e^{-2\pi
i\lambda_i\cdot\lambda_j}\qquad\hbox{if}\qquad\tau(0)=j,\eqno(3.18)$$
when $\g$ is simply laced and its roots are chosen to have length $\sqrt2$. It
is understood that $\lambda_0$ denotes zero.
\bigskip

 \ni{\bf 4. Fundamental Expectation Values}
\bigskip
\ni{\it 4.1 The Matrix $F$}
\medskip
The preceding work will help us solve two apparently unrelated problems at the
same time. The first is the evaluation of expectation values of the $r$
$\fiz$'s
with respect to the highest weight states $\rlj$ of the $r+1$ fundamental
representations:
$$F_{ji}=\llj\fiz\rlj\eqno(4.1)$$
as these play a crucial role in the expressions for soliton solutions. We shall
find the coefficient of proportionality between $F_{ji}$ and $F_{0i}$ which is
the \lq\lq vacuum expectation value" of $\fiz$. The second problem is the
determination of the action of the automorphism $\th$ on the $\fiz$. When
$\tau\in Aut\dg$, this will be useful for constructing the $\fiz$'s for the
nonsimply-laced Lie algebras via the folding of a simply-laced Lie algebra
\OTa, \GNOS .\medskip
Notice that (4.1) is independent of $z$ as only the zero (principal) mode of
$\fiz$ contributes. This means that we can equally well write
$$\fio=\sum_{j=0}^r\,h_jF_{ji}.\eqno(4.2)$$
The matrix $F$ is not square but there is a natural way of defining an extra
column with the label $0$ in terms of the level $x$:-
$$\hat F^0(z)=x\qquad\hbox{ so }\qquad F_{i0}=m_i.\eqno(4.3)$$
\medskip
Taking the $\rlj$ expectation value of the equation
$$[\ei,[\emi,\fio]]=\cases{|q(1)\cdot\gamma_i|^2\fio&$i\neq0$\cr
\qquad 0&$i=0$\cr}\eqno(4.4)$$
and substituting (2.16) and (4.2) yields
$$\sum_kC_{jk}F_{ki}=\cases{ |q(1)\cdot\gamma_i|^2F_{ji}&$
i\neq 0$\cr\qquad 0&$i=0$\cr}\eqno(4.5)$$
where
$$C_{jk}=m_jK_{jk}\eqno(4.6)$$
 and $K$ is the Cartan matrix of $\gh$. We see that the \lq\lq$i$"'th column of
the matrix $F$ furnishes an eigenvector of $C$ corresponding to the eigenvalue
$|q(1)\cdot\gamma_i|^2$ or zero. Notice that apart from the eigenvalue zero
these eigenvalues are proportional to the squared masses of the Toda
particles.
The matrix $C$, (4.6), is not symmetric, but it is symmetrisable, that is,
equivalent to its transpose. This means that there is a second scalar product
with respect to which $C$ is symmetric and its eigenvectors corresponding
to distinct
eigenvalues orthogonal. Unfortunately the eigenspaces of $C$ are degenerate in
general and the reason is simply the symmetry of $\gh$ which implies that
$D(\tau)$ commutes with $C$, by (3.4). That is, each eigenspace of $C$ carries
a representation of $Aut\dgh$ contained in $D$, (3.2). Because $\th$,
(3.5), respects the principal grades of $\gh$ we can take the zero mode
of (3.17) and insert (4.2) to find
$$\sum_kD_{jk}(\tau)F_{ki}=\epsilon(\tau,i)F_{ji},\qquad\tau\in
W_0(\g),\eqno(4.7)$$
where we define $\epsilon(\tau,0)=1$. Thus, the columns of $F$ are
 simultaneous eigenvectors of $C$ and $D(\tau)$.
 Because $Aut\dgh$ is, by definition, the symmetry group of $C$, and
because, by
 the result of
section 3, $W_0$ is its maximal abelian subgroup, we can consider
$C$ and $W_0$ as furnishing a complete set of commuting observables
whose simultaneous eigenvectors are the columns of $F$. These are
therefore orthogonal with respect to the second scalar product. Taking
them also to be normalised, the matrix $F$ is unitary with respect to
this product, and hence invertible. Considering now any $\tau\in
aut\dgh$, we find
$$\th(\fio)=\sum_j\fjo d_{ji}(\tau)\quad\hbox{ where }
d(\tau)=F^{-1}D(\tau)F.\eqno(4.8)$$
Thus $d$ is the matrix representing the action of the automorphism on
the $\fio$. It enjoys the following properties
$$d^{\dagger}(\tau)d(\tau)=1,\eqno(4.9a)$$
$$d_{i0}(\tau)=d_{0i}(\tau)=\delta_{i0},\eqno(4.9b)$$
$$d_{ij}(\tau)\hbox{ is diagonal, }\quad\tau\in W_0.\eqno(4.9c)$$
Unitarity, (4.9a), follows using the $Aut\dgh$ invariance of the second
scalar product. (4.9b) follows from unitarity and (3.6) while (4.9c)
expresses (3.17). When $\tau\in Aut\dg$, $D(\tau)$ also enjoys property
(4.9b). In fact we have verified case by case that, with a suitable
choice of labelling of the superscripts of the $\fiz$,
$$D(\tau)=d(\tau),\qquad\qquad\tau\in Aut\dg.\eqno(4.10)$$
Unfortunately we do not know a simple proof.
\medskip
\ni{\it 4.2 The Phases $\epsilon$}
\medskip
We are now in a position to be more specific about the
 phases $\epsilon(\tau,i)$ in (4.7) and (3.17). Since $W_0\cong Z(\g)$ is
abelian, its irreducible representations are all one dimensional. Evidently, by
(4.8), the phases $\epsilon(\tau,i) $ constitute the $r+1$ irreps of $W_0$
occurring in the decomposition of $D$, (3.2). There is a subset of these that
we
know precisely, namely those occurring in the representation of $W_0$ defined
by
$D$ acting on the tip points. The reason is that when  $W_0$ acts on these
tip points there are no fixed points so that it follows that this is the
regular
representation of $W_0$ and that each irreducible representation occurs
precisely
once. When $\g=G_2$ or $F_4$, $W_0$ is trivial but when $\g=B_r$ or $C_r$,
$W_0=Z_2$ and there are of course two irreps. It is more interesting when $\g$
is
simply laced. Then the tip points correspond to minimal weights (understanding
$\lambda_0=0$) and, if we choose the labelling appropriately, we have
$$\epsilon(\tau_i,j)=e^{-2\pi i\lambda_i\cdot\lambda_j}\quad\hbox{where}\quad
\tau_i(0)=i.\eqno(4.11)$$ Later on, in section 8, we shall present an argument
for the extension of (4.11) whereby
the label $j$ runs over all $r+1$ values, not just the $|W_0|$ values
corresponding to minimal weights.

\medskip
\ni{\it 4.3 Sample Calculations}
\medskip
The symmetry group of $\Delta(s\hat u(N))$ is the dihedral group $D_N$ while
$W_0$ is its cyclic subgroup $Z_N$. Labelling the vertices of
$\Delta (s\hat u(N))$ in the obvious consecutive manner $0,1,2,\dots N-1$, and
considering $W_0$ to be generated by $\tau_1(i)=i+1$, we find

$$F_{mn}=e^{2\pi imn/N}.\eqno(4.12)$$
Then $$d_{mn}(\tau_1)=\delta_{mn}e^{-2\pi im/N},$$
so that, if $\tau_j(i)=i+j=\tau_1^j(i)$,
$$d_{mn}(\tau_j)=\delta_{mi}e^{-2\pi imj/N}$$
in agreement with (4.11). Furthermore
$$c_{mn}\equiv(F^{-1}CF)_{mn}=\delta_{mn}\left(2sin{\pi m\over
N}\right)^2.$$  These yield the appropriate squared masses.
\medskip
The Lie algebra $D_4$ provides an interesting example as $Aut\Delta(\hat
D_4)\cong S_4$ while $W_0\equiv Z_2\otimes Z_2$ so, by (3.11),
$Aut\Delta(D_4)\cong S_3$. The eigenvalues of the matrix (4.6)
$$C=\left(\matrix{2&0&0&0&-1\cr0&2&0&0&-1\cr
0&0&2&0&-1\cr0&0&0&2&-1\cr-2&-2&-2&-2&4\cr}\right),$$
exhibits an unusually large degeneracy, being 0,2,2,2 and 6. Labelling the rows
and columns of this matrix 0,1,2,3 and 4 we have that the three non-trivial
elements of
the group $W_0$ take the following form when written in permutation notation:-
(01)(23), (02)(31) and (03)(12). Acting on the tip points 0,1,2 and 3 of
$\Delta(\hat D_4)$ each irreducible representation occurs precisely once
whereas the action on the central vertex 4 yields the scalar representation. We
then find that the matrix $F$ diagonalising $C$ and $W_0$ takes the form
$$F=\left(\matrix{1&1&1&1&1\cr1&1&-1&-1&1\cr1&-1&1&-1&1\cr1&-1&-1&1&1\cr
2&0&0&0&-4\cr}\right).\eqno(4.13)$$
Notice the order chosen for the columns 1,2 and 3 ensuring the entry 1 on the
diagonal given that the first line consists of unit entries. With this choice,
(4.11) again holds.
\bigskip

\ni{\bf 5.  Construction of $\fiz$ for Simply-Laced $\g$ at Unit Level}
\medskip
\ni{\it 5.1 The Principal Vertex Operator Construction }

\medskip
Our goal is to evaluate expectation values of products of $F$'s with
respect to the fundamental highest weight states $\rlj$, and hence
explicit soliton solutions. We have seen that for a single $F$ the
desired result is the solution to a problem involving finite matrices
of dimension $r+1$, whatever $\g$. The key to the more general problem
is the observation that sufficiently large powers of $\fiz$ always
vanish. Since the critical power involves the level and so is
representation dependent, our strategy will be to build up from the
irreducible representations with the simplest structure. These occur
when $\g$ is simply laced and at level $1$. The Frenkel-Kac-Segal
vertex operator construction \ref\FKS{ I.B. Frenkel and V.G. Kac, {\it
Inventiones
math.} {\bf 62} (1980) 23: G. Segal, {\it Commun. Math. Phys.} {\bf 80} (1981)
301} is familiar in this situation and there is a variant (with historical
priority) which expresses the $\fiz$ in terms of the principal Heisenberg
subalgebra, the $\em$. The highest weights $\Lambda_j$ of these irreducible
level
one representations, denoted $\rho_j$, correspond to the tip points of $\dgh$,
namely those related to the point $0$ by symmetries of $\dgh$ which can be
taken
to be unique elements of $W_0$. $\Lambda_0$ itself defines \lq\lq basic"
representation whose highest weight state is what physicists would
call the vacuum.
\medskip
Denoting \KKLW
$$\rho_j(\fiz)=F_{ji}exp\left(\sum_{N>0}{\gin z^N\hat E_{-N}\over
N}\right)exp\left(\sum_{N>0}{-\gin^*z^{-N}\en\over N}\right),\eqno(5.1)$$
where the sums extend over the positive affine exponents, we can
verify the correct commutation relations (2.18) with the $\en$ as well as the
correct expectation value in the highest weight state $\rlj$ by virtue
of (4.1). We also use the fact that $\rlj$ is the ground state of the Fock
space
built with the principal Heisenberg subalgebra:
$$\en\rlj=0\qquad N>0,\eqno(5.2)$$
and that this Fock space carries the irreducible representation $\rho_j$.
\medskip
\ni{\it 5.2 Operator Product Expansion of Two $F$'s}
\medskip
Using the familiar normal ordering procedures of string theory,
whereby the $\en$ with positive suffices are moved to the right of
those with negative suffices, we shall establish the crucial formula
$$\rho(\fiz)\rho(\fjz)=X_{i,j}(z,\zeta):\rho(\fiz)\rho(\fjz):\eqno(5.3a)$$
where
$$X_{i,j}(z,\zeta)=X_{j,i}(\zeta,z)=\prod_{p=1}^h\left(z-\omega^{-p}\zeta\right
)^{\sigma^p(\gamma_i)\cdot\gamma_j}.\eqno(5.3b)$$
Notice that the powers $\sigma^p(\gamma_i)\cdot\gamma_j$ in (5.3b) can only
take
the values $0,\pm1$ and $\pm2$ and we shall study the consequences of
this later.

 \medskip The proof of
(5.3) employs the standard techniques which immediately yield (5.3a) with
$$\xij=exp\left(-\sum_{N>0}{\gin^*\gjn\over N}({\zeta\over
z})^N\right),\eqno(5.4)$$ in which the sum converges when $|z|>|\zeta|$. The
sum
over the positive affine exponents $N=\nu+nh$, where $\nu=[N]$ is one of the
$r$
exponents of $\g$, can be written as a double sum over $n$ and $\nu$. Using the
identity for a subset of the logarithmic series
$$\sum_{n=0}^{\infty}{y^{nh+\nu}\over nh+\nu}=-{1\over
h}\sum_{p=1}^h
\omega^{p\nu}\hbox{ln}(1-y\omega^{-p}),$$
where $\omega$ is the primitive $h$'th root of unity, this becomes
$$\xij=exp\left(\sum_{p=1}^h\hbox{ln}(1-\omega^{-p}{\zeta\over
z})\Biggl\{{1\over h}\sum_{\nu}\omega^{p\nu}\gamma_i\cdot
q(\nu)^*\gamma_j\cdot q(\nu)\Biggr\}\right)$$
$$=\prod_{p=1}^h\left(1-\omega^{-p}{\zeta\over
z}\right)^{\sum_{\nu}\sigma^p\gamma_i\cdot q(\nu)^*\gamma_j\cdot q(\nu)/h},$$
where the sum over $\nu$ extends over the $r$ exponents of $\g$.
Because the $r$ eigenvectors of the Coxeter element $\sigma$ satisfy the
orthogonality property
$$q(\nu)^*\cdot q(\mu)=h\delta_{\mu,\nu},$$
we have the completeness relation
$$\sum_{\nu}q(\nu)q(\nu)^*=hI.$$
Inserting this in our previous expression for $\xij$ yields the
desired result (5.3), on realising that the sum of powers in (5.3b)
vanishes as $\sum_p\sigma^p=0$.
The symmetry of the factor $X$ cited in (5.3b) is very important
but not quite trivial to verify. Use is made of the fact just mentioned as
well as the identity
$\sum_{p=0}^{h-1}p\gamma_j\cdot\sigma^p\gamma_i=h\gamma_j\cdot\sigma^{-(1+c(i))
/2}\lambda_i\in
hZ\!\!\!Z$, where $c(i)$ denotes the \lq\lq colour" of the vertex $i$ in the
notation of \ref\FO{ A. Fring and D.I. Olive,  {\it Nucl. Phys.} {\bf 379B}
(1992) 429}
 and identity (2.7a) there is used.

\medskip
\ni{\it 5.3 Singularities of the OPE and Dorey's Fusing Rule}
\medskip
The fact that $\xij$ is symmetric (5.3b) means that we can calculate
commutators of the modes of the $F$'s in the familiar way, using
deformations of contour integrals. Here we
shall be content with studying the singularities of (5.3) which are
simply double and simple poles.
First note that by (5.3)
$$\fiz\fiz=0\eqno(5.5)$$
as the contribution of $p=h$ in the product vanishes and the other
factors are regular. This is nilpotency property appropriate to unit
level for simply laced algebras. The corresponding results
 at higher levels will be deduced from this.

Double poles can only occur in (5.3) at points $z=\omega^{-p}\zeta$
when the corresponding factor is raised to the power $-2$. this
requires
$$\sigma^p\gamma_i+\gamma_j=0$$ which is only possible when
$-\gamma_i$
and $\gamma_j$ lie on the same Coxeter orbit,
that is, $i$ and $j$ are conjugate. The precise value of $p$
in (5.6) was calculated in \FO. The coefficient of the
double pole is then a c-number while the coefficient of the associated
simple pole lies in the principal Heisenberg subalgebra.
\medskip
The remaining simple poles in (5.3b) can only occur at points
$z=\omega^{-p}\zeta$
and then only if the power of the responsible factor equals $-1$, that
is if
$\sigma^p\gamma_i+\gamma_j$ is a root. As each root  has a
unique expression $\sigma^q\gamma_k$, we have the condition
$$\sigma^p\gamma_i+\gamma_j=\sigma^q\gamma_k.\eqno(5.6)$$
To find the residue of this pole we collect the coefficient of $\hat
E_{-N}$ in the normal ordered product of (5.3b) and find
$$(\omega^{-p}\zeta)^N\gamma_i\cdot
q([N])+\zeta^N\gamma_jq([N])=(\omega^{-q}\zeta)^N\gamma_k\cdot
q([N])$$
using (5.6) and the fact that $q$ is an eigenfunction of $\sigma$. A
similar calculation applies to the coefficient of $\en$ and the
conclusion is that the residue is proportional to $\hat
F^k(\omega^{-q}\zeta)$.
This is the basis of our claim that the poles of the OPE of two $F$'s
are controlled by Dorey's fusing rule which therefore has a purely
Kac-Moody-Lie algebraic origin. Later we shall see that this statement
applies to all representations and that it has a consequence for the
\lq\lq fusing " of soliton solutions.
\medskip
\ni{\it 5.4 Arbitrary products of  $F$'s}
\medskip
Equation (5.3) for the product of two $F$'s can be extended to an
arbitrary product, again in the manner familiar from string theory:
$$\rho(\hat F^{i(1)}(z_1))\dots \rho(\hat F^{i(k)}(z_k))\dots\rho(\hat
F^{i(n)}(z_n))=\prod_{1\leq p<q\leq n}X_{i(p),i(q)}(z_p,z_q)$$
$$\times
:\rho(\hat F^{i(1)}(z_1))\dots\rho(\hat F^{i(k)}(z_k))\dots\rho(\hat
F^{i(k)}(z_n)):\eqno(5.7)$$
Initially the moduli of the arguments $z_k$ should decrease to the
right for convergence but as the right hand side of (5.7) is
meromorphic its domain of definition can be extended by analytic
continuation.
Because the same factors $X$ occur as before, (5.3), we can conclude
that the singularities arising are of the same type, that is, either due to
the occurrence of conjugate pairs or to the possibility of producing a third
soliton by Dorey's fusing rule. This conclusion is important as it provides the
basis for seeing that the products of Kac-Moody group elements occurring in the
soliton solutions are well defined and regular except at the special singular
points just mentioned. \medskip
Taking the expectation value of (5.7), with respect to the highest
weight state, $\rlj$, of the irrep $\rho$ and using (5.2) yields the matrix
element
$$\llj\hat F^{i(1)}(z_1)\dots\hat F^{i(k)}(z_k)\dots\hat
F^{i(n)}(z_n)\rlj=\prod_{1\leq p<p\leq
n}X_{i(p),i(q)}(z_p,z_q)\prod_{p=1}^nF_{ji(p)}.\eqno(5.8)$$

\ni{\bf 6. $\g$  Simply-Laced and Higher Levels of $\gh$}
\bigskip
\ni{\it 6.1 Nilpotency and OPE Properties of $\fiz$}
\medskip
When $\g$ is simply laced, the irreducible level $1$ representations
of $\gh$ are labelled by their $\g$ weights which are respectively $0$
and the fundamental minimal weights $\lambda_j$
$$|\Lambda_0>=|1,0>,\qquad\hbox{and}\qquad   \rlj=|1,\lambda_j>.$$
Each $\lambda_j$ defines a coset of the weight lattice of $\g$ with
respect to its root lattice. At higher levels $x>1$, we expect all
$\gh$ irreps with $\g$-weight in the same coset as $\lambda_j$ to
occur in the decomposition of the representation:
$$D^{(1,\lambda_j)}\otimes\overbrace{ D^{(1,0)}\otimes\dots D^{(1,0)}}
^{x-1\rm\,\, times}\eqno(6.1)$$
In general, it is not easy to perform the decomposition of (6.1) into
irreducibles but that is not necessary to establish the nilpotency and
operator product expansion properties of the $\fiz$ that we are seeking.
\medskip
Corresonding to (6.1) the construction of $\fiz$ at level two in terms
of the level one principal vertex operator construction is
$$\ffiz=\fiz\otimes 1+1\otimes\fiz, $$
and we see by (5.5), that
$$\ffiz^2=2\fiz\otimes\fiz.$$
So$$\ffiz^3=0\qquad\hbox{at level }2.$$
Repeating the construction at level $x$,
$$\ffiz=\fiz\otimes 1\otimes\dots1+1\otimes\fiz\otimes\dots 1+\dots
1\otimes1\otimes\dots\fiz, \eqno(6.2)$$
so$$\ffiz^x=x!\,\fiz\otimes\fiz\otimes\dots\fiz,\eqno(6.3)$$ we find the
previously announced nilpotency property
$$\ffiz^{x+1}=0.\eqno(6.4)$$
\medskip
We can also apply this style of argument to the operator product
expansion. Considering level $2$ for ease of writing, we have
$$\ffiz\ffjz=\fiz\fjz\otimes 1+1\otimes\fiz\fjz+\fiz\otimes
\fjz+\fjz\otimes\fiz.$$
Only the first two terms possess the singularities at $z$ in terms of
$\zeta$.
For example, the residue of the pole at $z=\omega^{-p}\zeta$ is
proportional to
$$\fkz\otimes 1+1\otimes\fkz=\ffkz.\eqno(6.5)$$
So Dorey's fusing rule still applies. The residue of the double pole
has doubled and so is proportional to the level.
\medskip
We conclude that at any level, $e^{Q\fiz}$ is well defined in the sense of
having finite matrix elements between any pair of states in the
representation space.  Thus, despite its superficial resemblance to a
vertex operator, it requires no normal ordering and so constitutes a bona-fide
Kac-Moody group element. As we have emphasised, the key point is that the
exponential series
terminates, by (6.4).

\bigskip
\ni{\it 6.2 Fundamental Expectation Values of $exp\,Q\fiz$}
\medskip
In order to evaluate the solution (2.27) for a single soliton of
species $i$, we seek the fundamental expectation values
$$\llj\epx e^{Q\fiz}\emx\rlj,\qquad j=0,1,2.\dots r.\eqno(6.6)$$
If $\lambda_j$ is minimal, so $m_j=1$, then, by (2.28), (2.29), (4.1)
and (6.4), this reduces to
$$\llj\epx(1+Q\fiz)\emx\rlj=e^{\mu^2x^+x^-}\left(1+F_{ji}QW(i)\right).\eqno(6.7
)$$
This suffices when $\g$ is of $A$-type but for the $D$ and $E$-type
simply laced algebras we may consider those $\Lambda_k$ corresponding
to vertices $k$ of $\dgh$ adjacent to a tip point j (with $m_j=1$).
Then $m_k=2$ and it is not difficult to show that
$$|\Lambda_k>={1\over\sqrt2}\left(\rlj\otimes
f_j\rlj-f_j\rlj\otimes\rlj\right)$$

$$={1\over\sqrt2}\left(\rlj\otimes\emi\rlj-\emi\rlj\otimes\rlj\right).\eqno(6.8
)$$

The expectation value of $\fiz$ with respect to $|\Lambda_k>$ is still
given by (4.1) whereas that of its square is found, using the
nilpotency properties (6.4), (6.3) and (6.8) to be given by
$$<\Lambda_k|\epx\fiz^2\emx|\Lambda_k>$$
$$=2\llj\epx\fiz\emx\rlj\llj\ei\epx\fiz\emx\emi\rlj$$
$$-2\llj\ei\epx\fiz\emx\rlj\llj\epx\fiz\emx\emi\rlj.$$
The matrix elements occurring here can be deduced by differentiating
(6.7) with respect to $x^+$, $x^-$ or both and the result after
substitution is simply
$$2e^{2\mu^2x^+x^-}\,F_{ji}^2\,W(i)^2.$$
 Thus, in this case, the
expectation value (6.6) is given by
 $$e^{2\mu^2x^+x^-}\left(1+F_{ki}QW(i)+F_{ji}^2Q^2W(i)^2\right).\eqno(6.9)$$

Notice that the dependence upon $z$ is wholly contained in the factor
$W(i)$ and that the result is clearly finite. Obviously this line of
argument
could be extended. The results so far are sufficient for the single
soliton solutions of affine $A_r$ and $D_4$, as we see in section 8.4.
\bigskip

\ni{\bf 7. $\g$ Non-Simply-Laced }
\medskip
We now seek to extend the construction of the $\fiz$ and the
verification of their properties to those untwisted affine Kac- Moody
algebras which are not simply laced. We shall exploit the existence
of  a natural embedding of each such algebra within a simply
laced one ($\g$ say) as the fixed subalgebra ($\gt$) of an
element $\tau$ of $Aut\dg$ lifted to $\g$ and so \lq\lq outer".
\medskip
 Before studying the effect on the $\em$ and $\fiz$ we shall recap
the basic ideas \OTa. We shall call $\tau$ \lq\lq direct" if within
each orbit of points given by its action on $\dg$ no two vertices
are linked. The Dynkin diagram of $\gt$, $\dgt$, is obtained by
\lq\lq folding" $\dg$, that is, by identifying  the vertices on each
separate orbit of $\tau$. The set of vertices contained in such an
orbit  containing vertex $i$ is denoted $<i>$. This will also be
used to label the vertices of $\dgt$. As $\tau$ preserves the point
$0$ the same folding procedure relates the extended Dynkin diagrams
of $\g$ and $\gt$. There is a precise correspondence between each of
these direct reductions and the list of simply laced Lie algebras.
This can be seen from the diagrams and is as follows:-
$$A_{2r-1}\rightarrow C_r, \quad E_6\rightarrow F_4,\quad
D_{r+1}\rightarrow B_r \hbox{ and } D_4\rightarrow G_2.\eqno(7.1)$$
\medskip Thinking either of $\g$ or $\gh$, we define
$$X_{<i>}=\sum_{i\in<i>}X_i\eqno(7.2)$$ where, as in (3.5), $X_i$
denotes $e_i$, $f_i$ or $h_i$. It is then easy to check the
commutation relations $$[h_{<i>},h_{<j>}]=0\eqno(7.3a)$$
$$[h_{<j>},e_{<i>}]=K_{<i><j>}e_{<i>},\eqno(7.3b)$$
$$[h_{<j>},f_{<i>}]=-K_{<i><j>}f_{<i>},\eqno(7.3c)$$
$$[e_{<i>},f_{<j>}]=\delta_{ij}h_{<i>},\eqno(7.3d)$$ where
$$K_{<i><j>}={1\over|<i>|}\sum_{i\in<i>,j\in<j>}K_{ij}=
\sum_{j\in<j>}K_{ij}.\eqno(7.4)$$
 The Serre relation (2.2) can also be checked. Once we have checked
that (7.4) indeed defines a Cartan matrix, we shall conclude that it
the Cartan matrix of $\gt$ since (7.3) constitute the defining
relations. \medskip Using the direct property, we see $K_{<i><i>}=2$
while, from the second
 expression (7.4) we see that when $<i>\neq<j>$, $K_{<i>,<j>}=0. -1,
-2,$ or $-3$, since the number of points in $<j>$, $|<j>|$, can only
equal $1, 2$ or $3$, according to the examples above.
\medskip
By the first version of (7.4) we see that
$${K_{<i><j>}\over K_{<j><i>}}={|<j>|\over|<i>|}.$$
In an obvious notation, this gives the ratio of squared root lengths
$\alpha_{<i>}^2/\alpha_{<j>}^2$. Since $\tau\in Aut\dg$, $|<0>|=1$ and
hence
$${\alpha_{<i>}^2\over\alpha_{<0>}^2}={1\over|<i>|}\eqno(7.5)$$

\medskip
By the definitions (2.4) together with the normalisation $m_0=n_0=1$,
we deduce
$$m_{<i>}=m_i, \quad n_{<i>}=|<i>|n_i \quad\hbox{and}\quad
h(\gt)=h(\g).\eqno(7.6)$$  The preservation of the Coxeter number $h$ in (7.6)
is very striking, but in fact the structure goes deeper. The principal $so(3)$
subalgebra of $\g$ plays a crucial role in the construction of the principal
Heisenberg subalgebra of $\gh$. We now show that it respected by
$\tau$ and hence survives as the principal $so(3)$ subalgebra of
$\gt$,
$$so(3)\subset\gt\subset\g.\eqno(7.7)$$
The reason is simply that the Weyl vector of $\g$ is $Aut\dg$
invariant. Also  the step operator for
the lowest root of both $\g$ and $\gt$ is identified. This together with (7.7),
again implies the equality of the Coxeter numbers.
\medskip
Turning to $\gh$ and $\bf{\hat g_{\tau}}$ we see that the principal
Heisenberg subalgebra of the latter is a subset of that of $\gh$. It
follows that the exponents of $\gt$ are a subset of the exponents of
$\g$ (with the same true automatically of the affine exponents).
Comparing with (3.9) and (3.13), we see that when $\tau\in Aut\dg$ is
direct and nontrivial, $\eta(\nu,\tau)=1$ if and only if $\nu$ is an
exponent of $\gt$. Looking at the examples (7.1), we note from tables that in
all
cases the common Coxeter number is even. Furthermore, with the exception of
$D_{2k}$, it is always precisely the even exponents of $\g$ which are deleted
in
order to obtain those of $\gt$. Hence for $\tau$ nontrivial and direct $\in
Aut\dg$ $$\th(\em)=(-1)^{M+1}\em\eqno(7.8)$$
with the exception of $\g=D_{2k}$. Since in this latter case, two
exponents equal $2k-1$, the principal  Heisenberg subalgebra at this
principal grade, $M=2k-1$, mod $4k-2$, must be two dimensional.
 As only one of these two exponents survive in the reduction to $B_{2k-1}$, we
must be able to choose a basis in the two-dimensional space so that

$$\th(\em)=\em,\qquad\th(\em^{\prime})=-\em^{\prime}.\eqno(7.9)$$
$D_4$ is exceptional in that $Aut\Delta({\bf D_r})$ is larger when $r=4$, being
$S_3$ the permutation group of the three tip vertices, which we label $1,2$ and
$3$. $D_4$ has exponents $1,3,3^{\prime}$ and $5$, while the reductions ${\bf
g_{(12)}}=B_3$ and ${\bf g_{(123)}}=G_2$ have exponents $1,3, 5$ and $1,3$
respectively. Choosing $\tau$  to be the permutation $(12)$, we can
define the basis at $M=6n+3$ to be as in (7.9). Since $S_3$ is represented in
this two-dimensional subspace and since the permutation $\sigma=(123)$ which is
responsible for the reduction to $G_2$ has no invariant subspace, this two
dimensional representation has to be the two dimensional irreducible
representation of $S_3$. Hence we calculate
$$\hat\sigma(\em)=-{1\over 2}\em-{\sqrt3\over2}\em^{\prime},\eqno(7.10a)$$
$$\hat\sigma(\em)^{\prime}={\sqrt3\over2}\em-{1\over2}\em^{\prime}.\eqno(7.10b)
$$
This completes what we have to say concerning the actions of the outer
automorphisms of affine untwisted Kac-Moody algebras on their principal
Heisenberg subalgebras.
\medskip
Given the choice of labelling whereby (4.10) holds, we can elevate
it to include all grades so that
$$\th(\fiz)=\hat F^j(z) D_{ji}(\tau)=\hat F^{\tau(i)}(z),\qquad\tau\in
Aut\dg\hbox{ and direct}.\eqno(7.11)$$
We now check this when $\tau$ has order two. On the basis of preceding
arguments we expect that
$$\th(\fiz)=\hat F^{\tau(i)}(z_{\tau})$$
and need only prove that $z_{\tau}=z$. Acting on the $\em$ $\fiz$ commutator
(2.18) with $\th$,
$$(-1)^{M+1}[\em,\hat F^{\tau(i)}(z_{\tau})]=z^M\gamma_i\cdot q(\mu)\hat
F^{\tau(i)}(z_{\tau}).$$
Now $\gamma_{\tau(i)}\cdot q(\mu)=(-1)^{M+1}\gamma_i\cdot q(\mu)$, using (2.23)
and the facts that $x_{\tau(i)}(\mu)=(-1)^{\mu+1}x_i(\mu)$, as shown in
FO and $\delta_{\tau(i)B}=\delta_{iB}$ as $\tau$ is direct.
Therefore $$[\em,\hat F^{\tau(i)}(z_{\tau})]=z^M\gamma_{\tau(i)}\cdot
q(\mu)\hat
F^{\tau(i)}(z_{\tau}).$$
Hence $z^M=z_{\tau}^M$ for all exponents $M$ of $\gh$. As this includes $M=1$
the result (7.11) follows.
 It then follows by the preceding discussion
that $$\hat F^{<i>}(z)=\sum_{i\in<i>}\fiz,\eqno(7.12)$$ where, again, $<i>$
labels an orbit of $\tau$. If $\tau$ is direct, we know that no two points $i$
and $j$, say, of $<i>$ are linked in $\dg$ and that, as a consequence,
$\gamma_i\pm\gamma_j$ cannot be a root. Hence, by the fusing rule,
$$[\fiz,\hat F^j(z)]=0,\eqno(7.13)$$ and we can repeat the arguments of section
6
to deduce that the highest non-vanishing power of $\hat F^{<i>}(z)$ is
$|<i>|x$,
remembering that the level, $x$, is preserved (3.6). By (7.5), this power
equals
$x\alpha_{<0>}^2/\alpha_{<i>}^2$ as claimed earlier. As this and lower powers
have finite matrix elements, we have completed our proof of the stated
properties of the powers of $\fiz$.

\bigskip

\ni{\bf 8. Soliton Solutions and Their Interpretation}
\medskip
\ni{\it 8.1 Topological quantum number and the phase $\epsilon$}
\medskip
The spatial jump of any soliton solution is conserved throughout time:
$${\partial\over\partial t}\Delta\phi=0\quad\hbox{ where
}\quad\Delta\phi\equiv\phi(t,x=\infty)-\phi(t,x=-\infty)\eqno(8.1)$$
This is because asymptotically the soliton solution must take values in
${2\pi i\over\beta}\Lambda_W(\g)$, where for the purposes of this section we
have assumed that $\g$ is simply laced with roots chosen to have length
$\sqrt2$. The expression (8.1) is called the topological quantum number but is
known to be difficult to calculate, even for single soliton solutions. The
reason is that it depends discontinuously on the phase of the parameter $Q$ in
(1.1) where $ln|Q|$ signifies the spatial coordinate of the soliton. However we
shall now see that, modulo $2\pi i\Lambda_R(\g)/\beta$, it is independent of
this
phase, depending only on the label $i$ in (1.1) in a way that we shall
determine.
\medskip
To do this we shall consider the quantities
$exp(-\beta\lambda_j\cdot\Delta\phi)$ which are easily calculable from
the general solution. Actually it is sufficient to consider only those
values of $j$ for which the fundamental weight $\lambda_j$ is minimal.
As the level of the corresponding representation with highest weight
state $\rlj$ is unity, the exponential (1.1) terminates after the
second term in the solution.
So
$$e^{-\beta\lambda_j\cdot\phi(x)}={1+Q\llj\fio\rlj W(i,z,x^{\mu})
\over1+Q\llo\fio\rlo W(i,z,x^{\mu})}$$
remembering (2.29). So, if the minus sign is taken in (2.30), so that a soliton
rather than an anti-soliton is
considered,
$$e^{-\beta\lambda_j\cdot\Delta\phi}={\llj\fio\rlj\over\llo\fio\rlo}.$$

The anti-soliton, coming from the alternative sign choice produces the inverse
factor. Now consider the unique element
of $W_0$ defined by $\tau_j(0)=j$. Putting $i=0$ in (3.7) and using
our freedom to choose
a phase to be unity, we have
$$\th_j\rlo=\rlj$$
which on insertion into the previous expression gives
$$e^{-\beta\lambda_j\cdot\Delta\phi}={\llo\th_j^{-1}\fio\th_j\rlo\over\llo\fio\
rlo}=\epsilon(\tau_j,i),\eqno(8.2)$$
using (3.17). Thus the phase $\epsilon$ acquires a direct
significance. At least when $\g$ is simply laced, explicit calculations
have led us to conjecture that
$$\epsilon(\tau_j,i)=e^{-2\pi
i\lambda_i\cdot\lambda_j}\eqno(8.3)$$
\medskip
Considering a solution describing several solitons of species
$i(1),i(2)\dots i(n)$, we find, by similar arguments that
$$e^{-\beta\lambda_j\cdot\Delta\phi}=\prod_{k=1}^n\epsilon(\tau_j,i(k)).
\eqno(8.4)$$
We shall find below that this taken together with the possibility of solitons
fusing gives support for (8.3).

\bigskip
\ni{\it 8.2 The \lq\lq Fusing" of Two solitons}
\medskip
We saw  that $$g=exp(Q_1\fif)exp(Q_2\fit)\eqno(8.5)$$ constitutes a
well-defined Kac-Moody group element unless $z_1$ and $z_2$ are related in the
exceptional ways discussed in section (5.3). Here we discuss the interpretation
of the singularity given by
$$z_1=\omega^{-p}z_2\quad\hbox{where $p$ is such that }\quad
\sigma^p\gamma_{i(1)}+\gamma_{i(2)}=\sigma^q\gamma_k,\eqno(8.6) $$
for some $q$ and $k$.
As the group element $g$ \lq\lq creates" two solitons of species $i(1)$ and
$i(2)$  whose rapidities, $ln|z_1|$ and $ln|z_2|$ coincide at the singular
point (8.6), we may suspect that some sort of bound state may be formed. We
shall confirm this and that the bound state is a third soliton, of
species $k$, given by Dorey's fusing rule.
\medskip
For simplicity, first consider the group element (8.5) evaluated at unit level
so
that $\fiz^2=0$, by (5.5). Also using (5.3), we find
$$g=1+Q_1\fif+Q_2\fit+Q_1Q_2X_{i(1),i(2)}(z_1,z_2):\fif\fit:.\eqno(8.7)$$
As the singular point (8.6) is approached, the factor $X$ in the last term
acquires a pole
while everything else remain finite if the constants $Q_i$ remain fixed.
Alternatively, we can suppose $Q_1$ and $Q_2$ tend to zero in such a way that
$Q_1Q_2X_{i(1),i(2)}$ remains finite. Then the limit of (8.7) is simply
$$g=1+Q^{\prime}\fkz=expQ^{\prime}\fkz,$$
again using (5.5) and letting $\zeta=z_2$ and $Q^{\prime}$ be a new constant.
Thus two solitons have
\lq\lq fused" to give a third with the selection rule given precisely by
Dorey's
fusing rule, originally formulated for the particle excitations of the theory
and now seen to extend to the solitons. It is not difficult to show that the
above limiting behaviour of the product of two Kac-Moody group elements
equalling a third is independent of the level.
\medskip
This result provides strong support for the validity of (8.3). Recall Braden's
result \ref\Brad{H.R. Braden, {\it J. Math. Phys.} {\bf A25} (1992) 215} that
if species $i(1)$ and $i(2)$ fuse to form $k$, then if the corresponding
fundamental representations are considered, $D^{\lambda_k}$ occurs in the the
Clebsch Gordon series of $D^{\lambda_{i(1)}}\otimes D^{\lambda_{i(2)}}$. This
implies that
 $$e^{-2\pi i\lambda_{i(1)}\cdot\lambda_j}e^{-2\pi
i\lambda_{i(2)}\cdot\lambda_j}=e^{-2\pi i\lambda_{k}\cdot\lambda_j}.$$
It follows that if the phases $\epsilon$ for two solitons are given by
(8.3) then so is that of any third soliton obtained by fusion, by (8.4).
\bigskip
\ni{\it 8.3 Breather Solutions}
\medskip
We have just seen how an analytic continuation of the parameters describing the
coordinates and rapidities of a solution describing two or more solitons can
lead to a new solution describing a lesser number of solitons. This phenomenon,
involving \lq\lq fusing", does not occur in the simplest affine Toda theory,
namely that with $\g=su(2)$, Sine Gordon theory, but there is a related
phenomenon that does. This is the occurrence of the \lq\lq breather" solution
\ref\Raj{R. Rajamaran, \lq\lq Solitons and Instantons", North Holland,
1982.}
which can be regarded as the bound state of soliton and antisoliton,
oscillating about their common centre of mass. Classically it possesses a
continuous mass spectrum extending from zero to the
soliton-antisoliton threshold. In the quantum theory, this continuous
spectrum is quantised, with the number of states depending on the
coupling constant $\beta$. The lowest mass such state (if it exists)
is identified with the quantum excitation of the original Sine-Gordon
field.
\medskip
We shall now see, by finding the corresponding Kac Moody \lq\lq group
element", that such solutions can be
formed of any soliton antisoliton pair in affine Toda theory, and that the
energy
and momentum can be evaluated by the techniques of our previous paper\OTU.

If $\delta_i$ denotes the phase of the complex number
$\gamma_i\cdot q(1)$, and
$$z_i=\epsilon_ie^{i\delta_i}e^{\eta_i},\qquad\epsilon_i=\pm1,$$
it was shown \OTU that, when $\eta_i$ is real, the contribution to
$\sqrt2$ times the light cone components $P^{\pm}$ of the momenta due
to a factor $expQ_i\hat F^i(z_i)$ in the group element $g(0)$, (2.27),
was $M_ie^{\pm\eta_i}$. $M_i$ is the mass of the soliton of species
$i$ and was calculated explicitly while $\eta_i$ can be interpreted as its
rapidity. In fact this argument holds good even if $e^{\eta_i}$ is
complex, providing that its real part is positive. Note that the
contribution to the momentum is independent of both the complex number
$Q_i$ and the sign $\epsilon_i$. Consider now a group element
$$g(0)=exp Q_1\hat F^1(e^{i\delta_1}e^{\eta_1})\,\,exp Q_2\hat
F^2(e^{i\delta_2}e^{\eta_2}).\eqno(8.8)$$
If $\eta_1$ and $\eta_2$ are real, this produces a two soliton
solution with momentum
$$\sqrt 2P^{\pm}=M_1e^{\pm\eta_1}+M_2e^{\pm\eta_2}.$$
However it is possible for this expression to remain real and positive
for a complex choice of rapidities $\eta_1$ and $\eta_2$, providing
$M_1=M_2=M$, say. Such a choice is given by
$$\eta_1=\eta_2^*=\eta+i\theta,\qquad -\pi/2<\theta<\pi/2,\eqno(8.9)$$
when
$$\sqrt2P^{\pm}=2Me^{\pm\eta}cos\,\theta.$$
Since the familiar $su(2)$, Sine-Gordon, breather arises this way,
(8.8) and (8.9) can be interpreted  as giving rise to a generalised breather
solution with mass
$2Mcos\,\theta$ and rapidity $\eta$.

Usually the equal mass condition is satisfied by choosing species $1$
and $2$ to be antiparticles of each other but in $D_4$ there is
another possibility, as, by triality, the $S_3$ symmetry of
$\Delta(D_4)$, there exist three solitons species of equal mass, as is
seen below in section 8.4. This raises the possibility of breather solutions
with non-zero topological charge which, at present, we are unable to
exclude. We anticipate that it will be important to evaluate the
higher conserved charges for such solutions.
\bigskip

\bigskip
\ni{\it 8.4 Sample Soliton Solutions}
\medskip

The affine $su(N)$ soliton solutions are particularly easy to evaluate
as all $N$ fundamental representations of $s\hat u(N)$ have level one.
The single soliton solution of species $I$ is given by
$$e^{-\beta\lambda_J\cdot\phi}={1+Qe^{2\pi
iIJ/N}W(I)\over1+QW(I)},\quad J=1,2,\dots N-1.$$
This follows from (2.27), (4.12) and (6.7). If we consider two
solitons of species $I$ and $J$, we find, using additionally (5.8), that
$e^{-\beta\lambda_J\cdot\phi}$ equals
$${1+e^{2\pi iI(1)J/N}W_1+e^{2\pi
iI(2)J/N}W_2+X_{I(1)I(2)}(z_1,z_2)e^{2\pi
i(I(1)+I(2))J/N}W_1W_2\over1+W_1+W_2+X_{I(1)I(2)}(z_1,z_2)W_1W_2},$$
where $J=1,2,\dots N-1$ and $W_q=Q(q)W(I(q))$ for short.

We can also immediately write down the $s\hat o(8)$ solitons. Using,
in addition, (4.13) and (6.9), we have for species $i=1,2$ and $3$,
corresponding to three of the four tip points of $\Delta(s\hat o(8))$,
$$e^{-\beta\lambda_j\cdot\phi}=\cases{1&$j=i$\cr {1-QW(i)\over 1+QW(i)}&$j\not=
 i,4$ \cr {1+Q^2W(i)^2\over(1+QW(i))^2}&$j=4$\cr},$$
while, for species $4$,
$$e^{-\beta\lambda_j\cdot\phi}=\cases{1&j=1,2,3\cr
{1-4QW(4)+Q^2W(4)^2\over(1+QW(4))^2}&j=4\cr}.$$
\medskip
Using the additional results of section 7, concerning non simply laced
algebras, we can construct the two species of affine $G_2$ soliton solution.
By (7.12), these are created by
$$\hat F^{<1>}=\hat F^{<2>}=\hat F^{<3>}=\hat F^1+\hat F^2+\hat
F^3\quad\hbox{and}\quad\hat F^{<4>}=\hat F^4,$$
in terms of the quantities for $\hat D_4$. We have retained the
convention for labeling the vertices of $\Delta(\hat D_4)$ $0,1,2,3$ and
$4$. It follows from their defining properties that the fundamental
highest weight states of $\hat G_2$ are
$$|\Lambda_{<0>}>=|\Lambda_0>,\qquad |\Lambda_{<1>}>=|\Lambda_1>\hbox{
or }|\Lambda_2>\hbox{ or }|\Lambda_3>.$$
So the $\hat F^{<4>}$ solution for affine $G_2$ is given by the same
expression as the affine $D_4$ solution above. For the other soliton
solution, we have, for $j=0$ or $1$
$$<\Lambda_{<j>}|\hat
F^{<1>}(z)|\Lambda_{<j>}>=F_{j1}+F_{j2}+F_{j3},$$
$$<\Lambda_{<j>}|\hat
F^{<1>}(z)^2|\Lambda_{<j>}>=2X_{12}(z,z)(F_{j1}F_{j2}+F_{j2}F_{j3}+F_{j3}F_{j1}
),$$
$$<\Lambda_{<j>}|\hat
F^{<1>}(z)^3|\Lambda_{<j>}>=6X_{12}(z,z)^3F_{j1}F_{j2}F_{j3},$$
where the $\hat D_4$ matrix $F_{ji}$ (4.13) enters the right hand
side and (5.7) has been used. As $h=6$ for $D_4$ or $G_2$, we can
evaluate $X_{12}(z,z)=1/3$ using (5.3b). Inserting these numerical
values, (2.27) yields
$$e^{-\beta\lambda_{<1>}\cdot\phi}={1-W-{1\over3}W^2+{1\over27}W^3\over1+3W+W^2
+{1\over27}W^3},$$
where $W=QW_{<1>}$. The other component of this solution is more
difficult to evaluate since powers of $\hat F$ up to six survive.
Combining the previous methods, we find
$$e^{-\beta\lambda_{<4>}\cdot\phi}={1+3W^2-{16\over27}W^3+{1\over3}W^4+{1\over2
7^2}W^6\over\left(1+3W+W^2+{1\over27}W^3\right)^2}.$$

All the single soliton solutions for the affine untwisted Toda
theories have recently been worked out explicitly using Hirota's
method \Ar. The results so obtained for the enumeration of soliton species
and for the mass formulae seem to agree with the results of the general
arguments of \OTU rather than with \MM whose results were incomplete. It
is interesting to see that although  the general features of
 our approach eventually emerge, this Hirota method
 appears comparatively cumbersome.

\bigskip

\ni{\bf 9. Discussion}
\medskip
We have succeeded in our main aim of establishing the conditions under
which products of Kac-Moody group elements of the form (1.1) make
sense. The key result is the proof of the vanishing, in a representation of
level $x$,
of all powers of $\fiz$ exceeding $2x/\gamma_i^2$. In the course of
the work some new structural features  have emerged as well as new questions
such as the general
proof of (8.3).
\medskip
The extension of Dorey's fusing rule to solitons, (at least when $\g$ is
simply laced), strengthens the evidence for a duality symmetry between the
particles which are the quantum excitations of the elementary fields and the
solitons. An intriguing  difference concerns the fact that the solitons
possess an internal degree of freedom due to the different values of the
topological charge possible for a soliton of a given species. It is therefore
urgent to clarify the structure of this spectrum but  old questions persist
such as
the dependence of the topological charge (8.1) upon the phase of the constant
$Q$
in (1.1). But we have been able to show that,   modulo $2\pi
i\Lambda_R(\g)/\beta$,
it is independent of $Q$. Although we have seen how the breathers of familiar
type fit
naturally within our formalism, we have not been able to exclude breathers of a
new kind.

 \medskip There are many ways in which we have good reason to
believe our arguments can be extended. One concerns the affine Toda theories
associated with twisted affine Kac-Moody algebras (rather than just the
untwisted ones we have considered). Another concerns the non-Abelian Toda
theories corresponding to non-principal embeddings of $SO(3)$ in $\g$.
\medskip
\noindent{\bf Acknowledgements}
\medskip

DIO wishes to thank Peter Johnson and Marco Kneipp for discussions. JWRU thanks
SERC for the
award of a studentship while JWRU and DIO thank the Isaac Newton Institute for
hospitality
during the early stages of this work. NT was partially funded by NSF contract
No PHY90-21984,
the Alfred Sloan Foundation and the David and Lucile Packard Foundation.

\listrefs
\bye